\documentclass{aastex631}

\usepackage{amsmath}
\usepackage{multirow}

\begin{document}

\title{The Strength and Shapes of Contact Binary Objects}

\correspondingauthor{Alex J Meyer}
\email{alex.meyer@colorado.edu}

\author{Alex J. Meyer}
\affiliation{Smead Department of Aerospace Engineering Sciences, University of Colorado Boulder, 3775 Discover Dr, Boulder, CO 80303, USA}

\author{Daniel J. Scheeres}
\affiliation{Smead Department of Aerospace Engineering Sciences, University of Colorado Boulder, 3775 Discover Dr, Boulder, CO 80303, USA}



\begin{abstract}

While contact binary objects are common in the solar system, their formation mechanism is unclear. In this work we examine several contact binaries and calculate the necessary strength parameters that allow the two lobes to merge without the smaller of the two being gravitationally destroyed by the larger. We find a small but non-zero amount of cohesion or a large friction angle is required for the smaller lobe to survive the merging process, consistent with observations. This means it is possible for two previously separated rubble piles to experience a collapse of their mutual orbit and form a contact binary. The necessary strength required to survive this merger depends on the relative size, shape, and density of the body, with prolate shapes requiring more cohesion than oblate shapes. 

\end{abstract}



\section{Introduction} \label{sec:intro}
Contact binary objects, bodies which are clearly defined by two distinct lobes, are abundant in the solar system. Recent estimates place the occurrence of contact binaries for Near-Earth asteroids (NEAs) with a diameter larger than 200 m at around 30\% \citep{virkki2022arecibo}. Contact binaries are also common in the populations of comets \citep{nesvorny2018bi}, Kuiper-Belt objects (KBOs) \citep{sheppard2004extreme}, and Jupiter Trojans \citep{mann2007fraction}. In this work we find contact binaries can be made up of two rubble-pile bodies that merged after a collapse of their mutual orbit, provided the rubble piles have large friction angles or a small amount of cohesion, consistent with values found in the literature. This means monolithic constituents within the rubble piles are not required for the system to survive the merger process.

The specific formation process of contact binaries is unclear. \cite{bagatin2020gravitational} hypothesize these objects are made via gravitational re-accumulation following a catastrophic disruption of a progenitor body. In this model, asteroid shapes are dominated by collisional processes. This has the advantage of applicability throughout the solar system. 

Another hypothesis was presented by \cite{jacobson2011dynamics}, who argue contact binaries form as a result of a collapsed binary system. This requires rotational spin-up of the progenitor via the YORP effect to the point of mass loss. The resulting satellite then drifts inward due to the Binary YORP (BYORP) effect \citep{cuk2005effects,cuk2010orbital}, until the two asteroids merge in a low speed collision.

\cite{nesvorny2019trans} demonstrate that KBO contact binaries are possibly formed by the streaming instability, which may also be applied to the formation of the Jupiter Trojans \citep{morbidelli2005chaotic}. In this approach, rather than formation from a single progenitor body the resultant contact binary is formed from the gravitational collapse within a protoplanetary disk \citep{youdin2005streaming}.

Regardless of the specific formation mechanism, contact binaries appear to be two separate bodies which have merged, requiring them to have been gravitationally bound in the past. This means the smaller body must have been able to withstand the gravitational tides from the larger body during the merger. In this work, we study several contact binary objects and calculate the necessary physical properties of the smaller lobe for it to withstand the larger lobe's gravitational tides in order to maintain its shape as it merges. This carries the implicit assumption that the smaller secondary will fail before the larger primary, which we will motivate below. These constraints elucidate the feasibility of this model as a method of forming contact binaries, as well as providing additional constraints on the population of small solar system bodies.

In this paper, we first present the contact binaries used in our analysis in Section \ref{sec:shapes}. We then outline the failure criteria used to calculate the minimum strength, and report these results in Section \ref{sec:failure}. In this section we also present a discussion of the simple sphere-sphere model of contact binaries. In Section \ref{sec:population} we compare the population of contact binaries to that of binary asteroids and asteroid pairs. Finally, we discuss the implications of this work in Section \ref{sec:discussion}.

\section{Contact Binaries} \label{sec:shapes}

For our study we analyze contact binaries for which radar- or spacecraft-derived shape models exist in the literature. These are all the solar system bodies which have been identified as contact binaries and for which a shape model exist. This gives us 11 contact binaries: 7 asteroids (25143) Itokawa, (4179) Toutatis, (8567) 1996HW1, (85990) 1999JV6, (4769) Castalia, (4486) Mithra, and (2063) Bacchus; 3 comets 67P/Churyumov-Gerasimenko, 103P/Hartley, and 8P/Tuttle; and 1 Kuiper belt object (486958) Arrokoth. For brevity, henceforth we will leave off the minor planet numbers when referring to these objects, and we will abbreviate comet 67P/Churyumov-Gerasimenko as 67P/C-G. 

In these systems, we will refer to the larger lobe of the contact binary as the primary, and the smaller lobe as the secondary. For our analysis we will need to fit ellipsoids to both the primary and the secondary in each system. Using the published polyhedral shape models, we do this using a simple linear least-squares fitting algorithm implemented in Matlab \citep{fitellipsoid}. A few representative cases are illustrated in Appendix \ref{apdx:A}, showing a comparison between the polyhedral shape models and our ellipsoid fitting for some contact binary asteroids.

In our model, before the primary and secondary merge to form a contact binary, they are orbiting one another very closely. In order for the secondary to survive the merger process, the Roche limit must be smaller than the sum of the radii of the primary and secondary. Because we will take an averaged approach to the stresses, we are not enforcing a condition of no mass loss from the secondary, but only the secondary's resistance to complete structural failure when it comes into contact with the primary. Thus, we will calculate the minimum bulk structural properties of the secondaries of these contact binaries that allow them to survive the merger process.

\section{Failure Criteria} \label{sec:failure}

We use the Drucker-Prager strength model to identify the failure criteria, which can be applied to rubble-pile bodies. In their work, \cite{holsapple2006tidal,holsapple2008tidal} apply this model to an ellipsoidal body in proximity to a larger, spherical gravitational body, and we use this approach here. The failure criteria is
\begin{equation}
    \sqrt{J_2}=k-sI_1
\end{equation}
where $k$ refers to the bulk cohesive strength of the body and $s$ is the shear strength which depends on the material friction angle $\phi$. $I_1$ is the first invariant of the Cauchy stress tensor, and $J_2$ is the second invariant of the deviatoric stress tensor. For a detailed discussion on this failure criteria and how it's applied in this work, see Appendix \ref{apdx:B}.  

This approach makes the assumption that the larger gravitational body is spherical. However, as we will see, the primaries in the contact binaries used in this investigation are non-spherical. While this difference in shape will have some effect on the required material properties, we are using this approach as an order of magnitude study. We leave the consideration of a non-spherical primary to future study.

\begin{figure}[ht!]
   \centering
   \includegraphics[width = 3in]{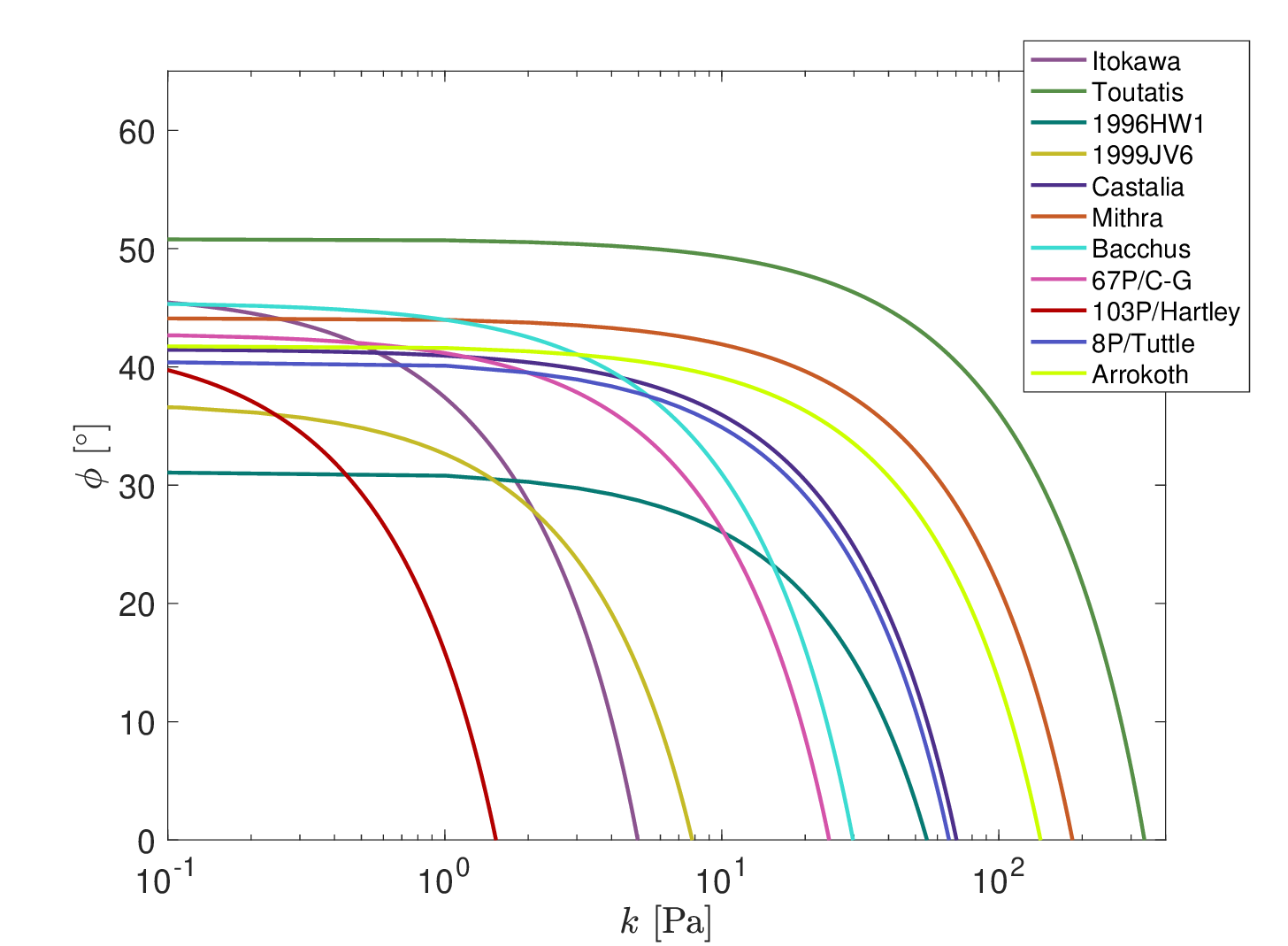} 
   \caption{The friction angle as a function of cohesion for our 10 contact binaries. These curves represent the critical combinations of friction angle and bulk cohesion that result in structural failure of the secondary lobe of the contact binary.}
   \label{fig:friction_cohesion}
\end{figure}

Failure of the secondary is counteracted by both its cohesive and shear strength, which we respectively parameterize by cohesion $k$ and friction angle $\phi$. For each contact binary in our investigation, we plot curves of $\phi$ versus $k$ in Figure \ref{fig:friction_cohesion}. This demonstrates the relationship between these parameters: as friction angle increases, less cohesion is required, and vice versa. For large enough friction angles, no cohesion is required, and for large enough cohesion, no friction is required. We note these curves represent the failure criteria, thus any combination of friction angle and cohesion outside these curves is also permissible.

We present a detailed outline of the parameters used in this analysis in Table \ref{tab:parameters}. This includes results of the ellipsoidal fitting to both the primary and secondary. The primary ellipsoid has a volume-equivalent diameter of $D_1$, and ellipsoidal semi-axes $a_1>b_1>c_1$. The secondary ellipsoid has a volume-equivalent diameter of $D_2$ and ellipsoidal semi-axes $a_2>b_2>c_2$. We also report the densities of the primary and secondary, $\rho_1$ and $\rho_2$, respectively. When density estimates exist for the bodies we use those published values. If no estimate exists, we assume a density of $2.5\mathrm{ g/cm}^3$ for asteroids and $0.6\mathrm{ g/cm}^3$ for comets. For all bodies except Itokawa, we assume the primary and secondary have identical densities. For Itokawa, we use primary and secondary density estimates from \cite{kanamaru2019density}.

In Table \ref{tab:parameters}, we also report the minimum cohesion required assuming a friction angle of $30^\circ$. Across all 11 contact binaries, this minimum cohesion is on the order of 1-100 Pa. This is just an order of magnitude study, and thus we do not have formal uncertainties on these values. However, we see a small but non-zero amount of cohesion is required for all bodies, unless the friction angle becomes large.

\begin{table}
\caption{The results of our ellipsoidal fitting to the contact binary shape models. We report the rotation period, the diameter and shape ratios of the primary, the diameter and shape ratios of the secondary, the densities, and the minimum cohesion required to form a contact binary assuming a $30^\circ$ friction angle.}
\label{tab:parameters}
\centering
\begin{tabular}{ |c||c|c|c|c|c|c|c|c|c|c|c|  }
 \hline
 \multicolumn{1}{|c||}{ } & \multicolumn{5}{|c|}{Primary} & \multicolumn{4}{|c|}{Secondary} & \multicolumn{2}{|c|}{ } \\
 \hline
Name & P [hr] & $D_1$ [km] & $a_1/b_1$ & $b_1/c_1$ & $\rho_1$ [g/cm$^3$] & $D_2$ [km] & $a_2/b_2$ & $b_2/c_2$ & $\rho_2$ [g/cm$^3$] & min $k$ [Pa] & Reference \\
\hline
Itokawa & 12.13 & 0.31 & 1.82 & 1.29 & 1.93 & 0.20 & 1.35 & 1.25 & 2.45 &  1.85 & 1, 2 \\
Toutatis & 177 & 2.36 & 1.54 & 1.20 & 2.50$^*$ & 1.41 & 1.46 & 1.21 & 2.50$^*$ & 143 & 3, 4 \\
1996HW1 & 8.77 & 1.77 & 1.57 & 1.08 & 1.73 & 1.41 & 1.10 & 1.05 & 1.73$^*$ & 3.01 & 5 \\
1999JV6 & 6.54 & 0.41 & 1.76 & 1.03 & 2.50$^*$ & 0.32 & 1.19 & 1.06 & 2.50$^*$ & 1.61 & 6 \\
Castalia & 4.10 & 0.94 & 1.08 & 1.13 & 2.50$^*$ & 0.84 & 1.12 & 1.42 & 2.50$^*$ & 20.8 & 7 \\
Mithra & 67.5 & 1.44 & 1.20 & 1.12 & 2.50$^*$ & 1.25 & 1.31 & 1.19 & 2.50$^*$ & 63.0 & 8 \\
Bacchus & 14.9 & 0.58 & 1.64 & 1.01 & 2.50$^*$ & 0.49 & 1.43 & 1.08 & 2.50$^*$ & 10.7 & 9 \\
67P/C-G & 12.4 & 3.13 & 1.29 & 1.19 & 0.53 & 2.26 & 1.22 & 1.21 & 0.53$^*$ & 7.81 & 10 \\
103P/Hartley & 18.3 & 1.03 & 1.83 & 1.15 & 0.40 & 0.77 & 1.31 & 1.06 & 0.40$^*$ & 0.48 & 11 \\
8P/Tuttle & 11.39 & 4.68 & 1.38 & 1.31 & 0.60$^*$ & 3.50 & 1.31 & 1.00 & 0.60$^*$ & 19.0 & 12 \\
Arrokoth & 15.94 & 14.0 & 1.12 & 3.09 & 235 & 12.6 & 1.12 & 1.44 & 235$^*$ & 43.0 & 13, 14 \\ 
 \hline
\multicolumn{12}{>{\raggedright}p{\textwidth}}{$^\mathrm{*}$For systems lacking data, density values are assumed using $2.5$ g/cm$^3$ for asteroids and $0.6$ g/cm$^3$ for comets. The contact binaries are assumed to have the same bulk density in both lobes unless separate measurements exist. $^\mathrm{1}$\cite{gaskell2008} $^\mathrm{2}$\cite{kanamaru2019density} $^\mathrm{3}$\cite{hudson2003high} $^\mathrm{4}$\cite{hudson1995shape} $^\mathrm{5}$\cite{magri2011radar} $^\mathrm{6}$\cite{rozek2019shape} $^\mathrm{7}$\cite{hudson1994shape} $^\mathrm{8}$\cite{brozovic2010radar} $^\mathrm{9}$\cite{benner2000asteroid} $^\mathrm{10}$\cite{jorda2016global} $^\mathrm{11}$\cite{thomas2013shape} $^\mathrm{12}$\cite{harmon2010radar} $^\mathrm{13}$\cite{stern2019initial} $^\mathrm{14}$\cite{keane2022geophysical}} 
\end{tabular}
\end{table}

To compare these results, we non-dimensionalize the systems using the primary's radius and mass as the length and mass units, respectively. The time unit is the orbit period at the primary's radius. Thus, the normalized cohesion is defined as
\begin{equation}
    \kappa = \frac{k}{\rho_1^2R_1^2G}.
\end{equation}

As we will see, the shape of the secondary is also an important consideration. We parameterize the ellipsoidal shapes using the axis ratios: $a_2/b_2$ and $b_2/c_2$. 

\begin{figure}[ht!]
    \gridline{
            \fig{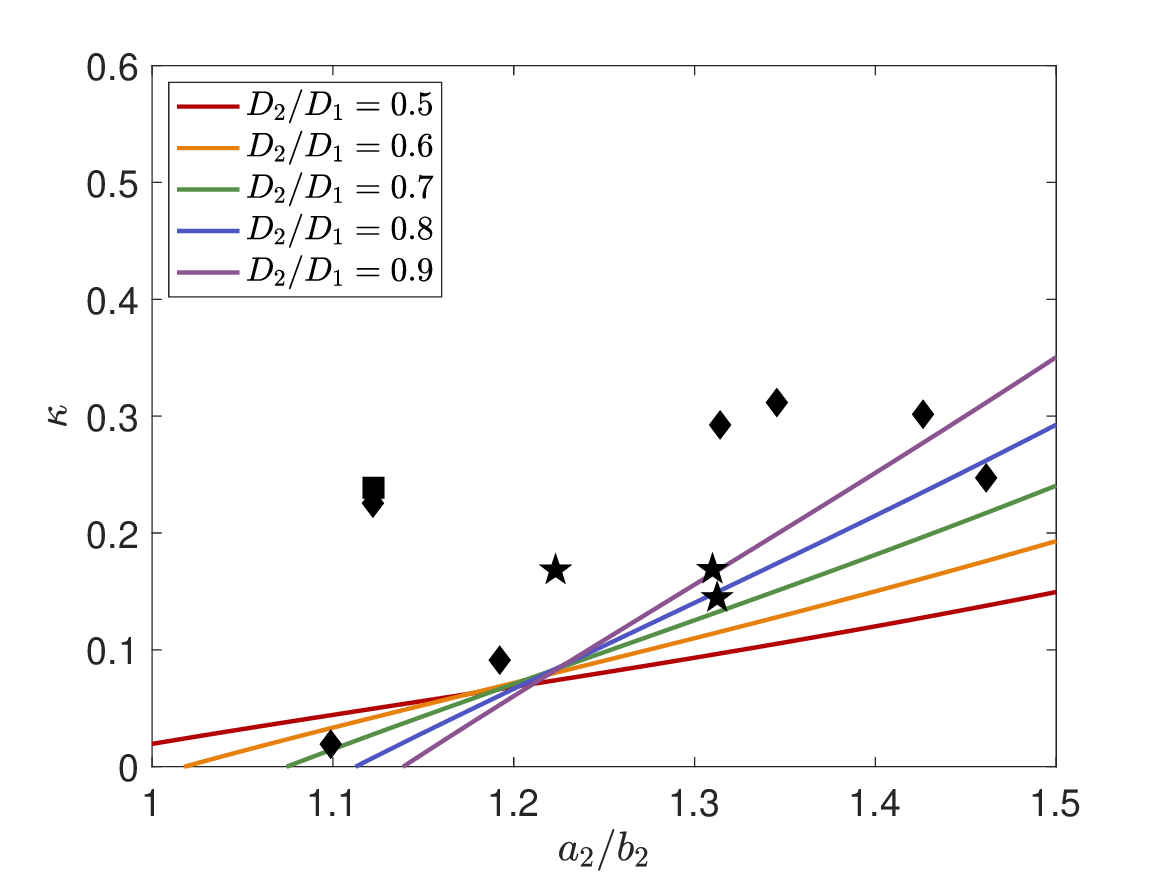}{0.45\textwidth}{(a) Cohesion as a function of prolateness.}
            \fig{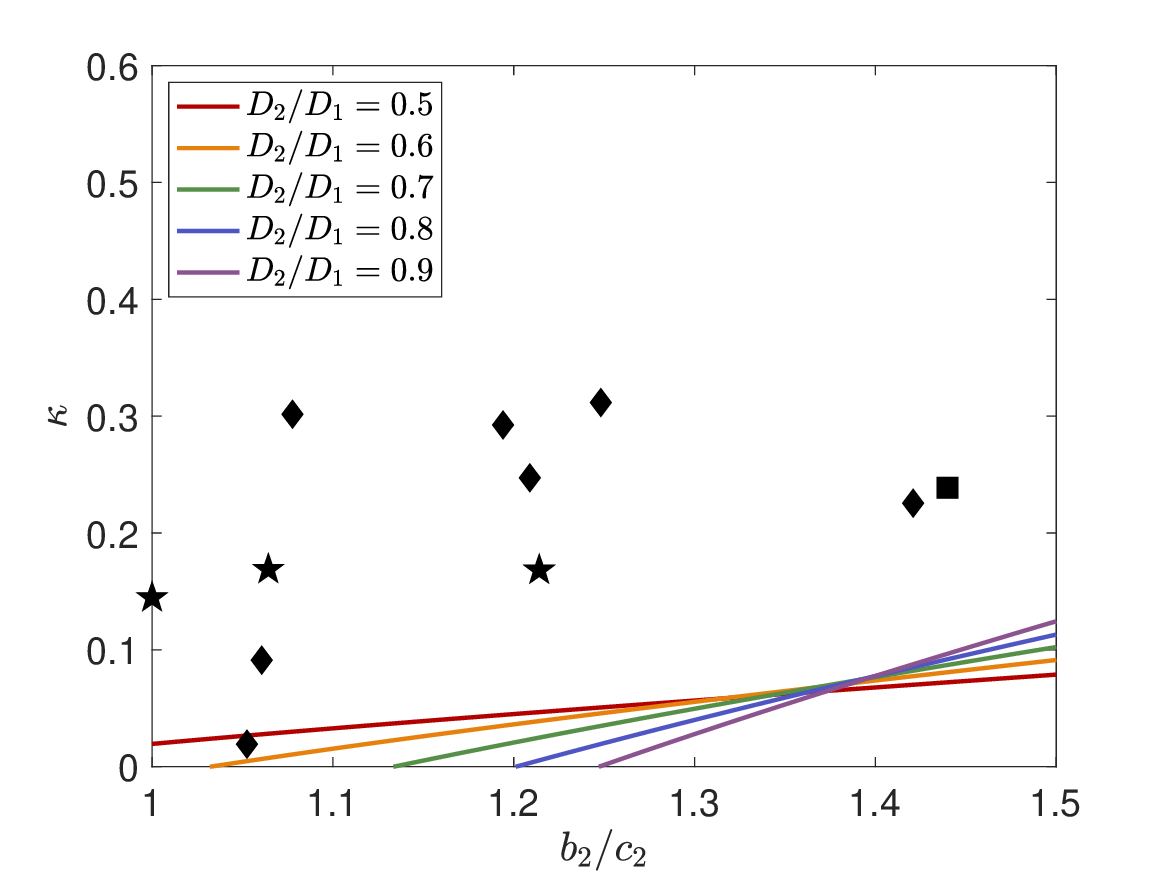}{0.45\textwidth}{(b) Cohesion as a function of oblateness.}
            }
    \gridline{
            \fig{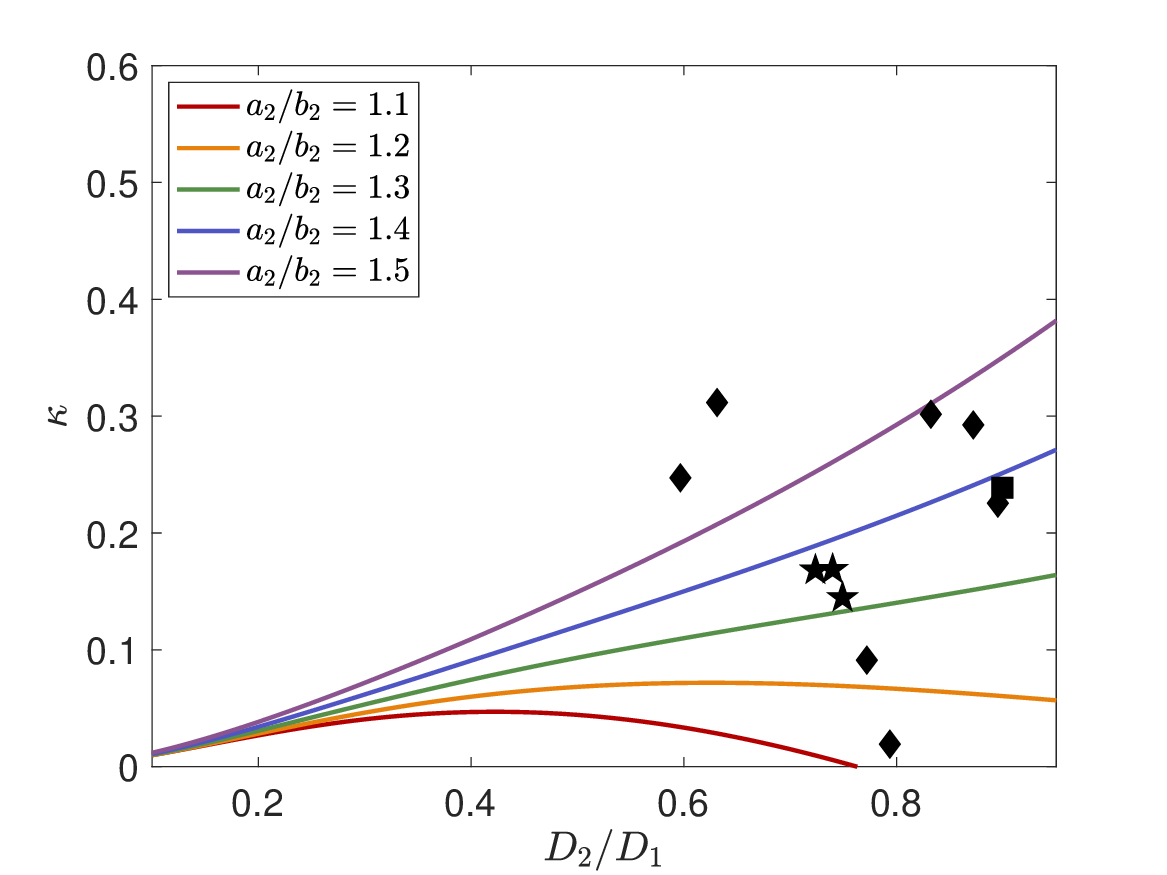}{0.45\textwidth}{(c) Cohesion as a function of size ratio for prolate bodies.}
            \fig{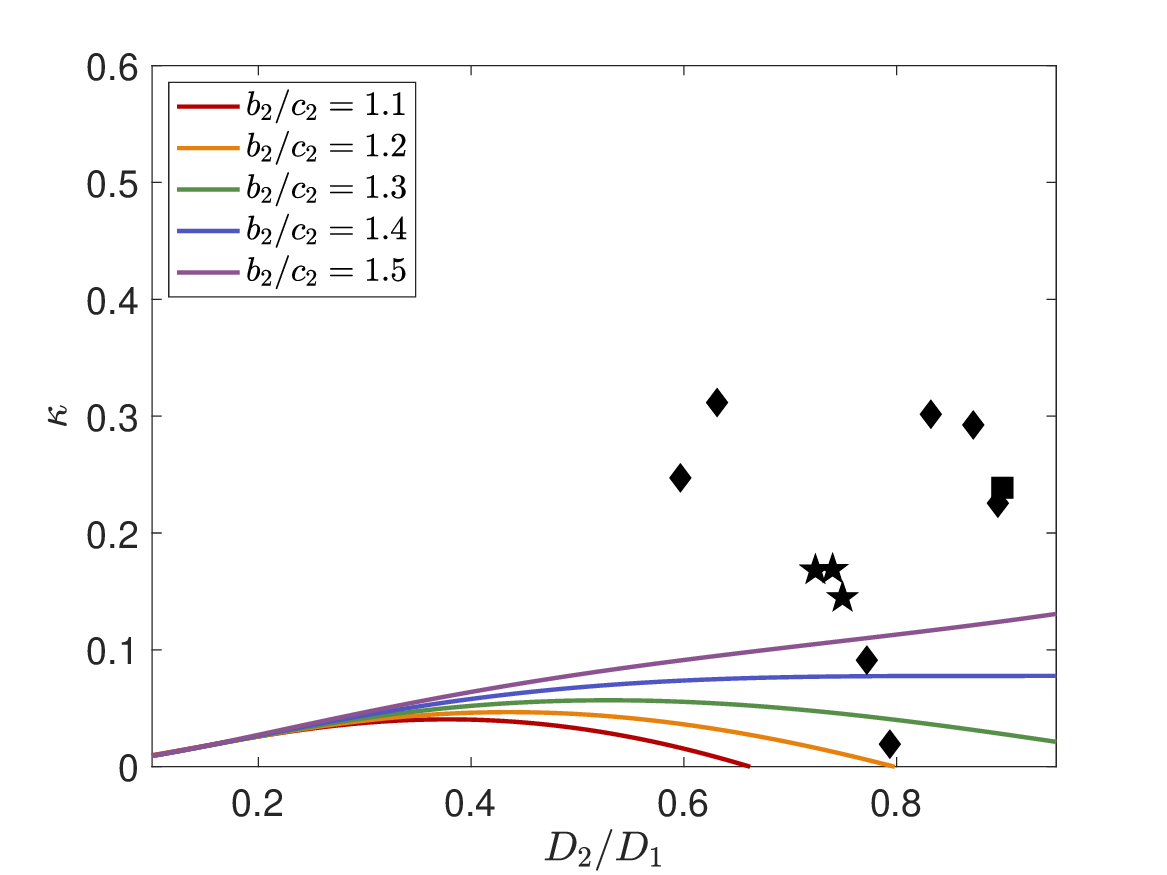}{0.45\textwidth}{(d) Cohesion as a function of size ratio for oblate bodies.}
            }
    \caption{Plots showing the minimum normalized cohesion as a function of: (a)  the prolateness of the secondary lobe at different size ratios; (b) the oblateness of the secondary lobe at different size ratios; (c) the size ratio for prolate secondary lobes; (d) the size ratio for prolate secondary lobes. Our contact binaries are plotted as data points, with asteroids shown as diamonds, comets as stars, and KBOs as squares.}
    \label{fig:cohesion_general}
\end{figure}

Assuming a typical friction angle of $30^\circ$ and equal primary and secondary densities, we investigate the role of secondary-to-primary size ratio and prolateness or oblateness in Figure \ref{fig:cohesion_general}. In investigating prolateness, we set $b_2/c_2=1$ and vary $a_2/b_2$, and for investigating oblateness we set $a_2/b_2=1$ and vary $b_2/c_2$. We also plot our contact binaries on these plots: asteroids are plotted as diamonds, comets as stars, and the KBO as a square. This reveals a complicated relationship between cohesion, size ratio, and asphericity.

Generally, we see increasing $a_2/b_2$ or $b_2/c_2$ results in an increased minimum cohesion for a given size ratio. This effect is more pronounced for prolate shapes (increasing $a_2/b_2$), and less important for oblate shapes (increasing $b_2/c_2$). However, in both cases we see a switch in the effect of the size ratio. For nearly spherical secondary lobes with small $a_2/b_2$ or $b_2/c_2$, increasing size ratio requires less cohesion. The opposite is true for very oblate or prolate shapes, where larger size ratios require more cohesion. This switch occurs around $a_2/b_2\sim1.2$ for prolate shapes and around $b_2/c_2\sim1.38$ for oblate shapes. However, this only holds for size ratios $D_2/D_1\gtrsim0.4$. For small size ratios, the required cohesion always increases with size ratio.

\begin{figure}[ht!]
   \gridline{
            \fig{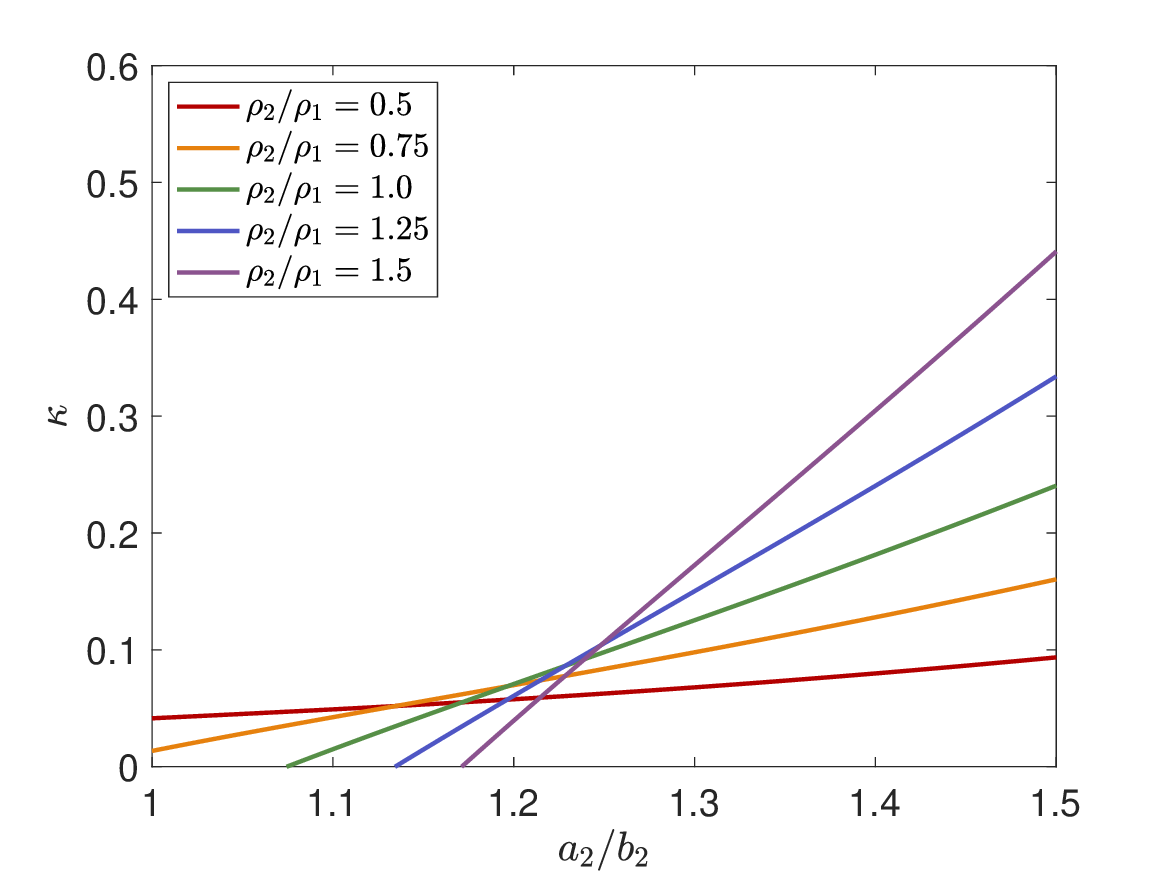}{0.45\textwidth}{(a) Cohesion as a function of prolateness for density ratios.}
            \fig{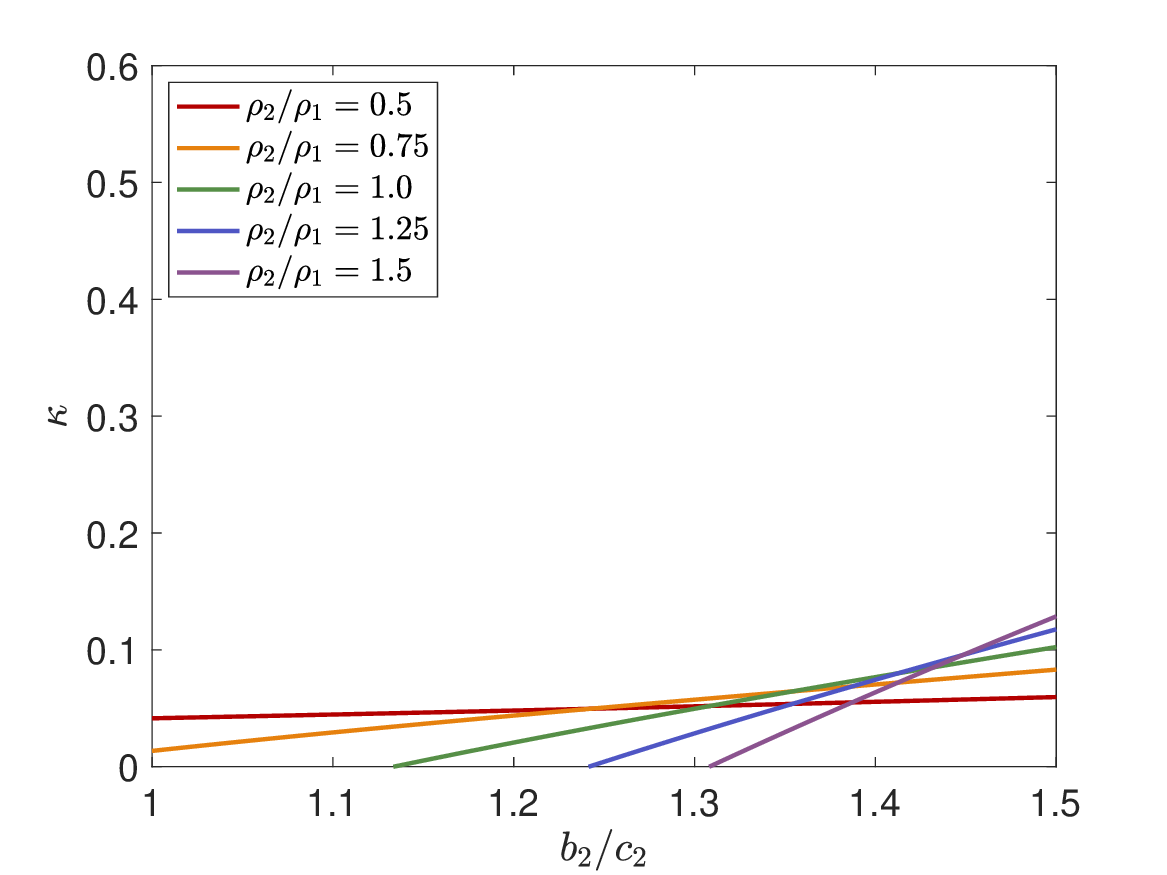}{0.45\textwidth}{(b) Cohesion as a function of oblateness for density ratios.}
            } 
   \caption{The minimum normalized cohesion as a function of (a) prolateness and (b) oblateness of the secondary lobe for different values of the density ratio. The curves assume a size ratio of $D_2/D_1=0.7$.}
   \label{fig:cohesion_density}
\end{figure}

Up to this point, we have assumed the primary and secondary have identical bulk densities. We next examine the role played by the density ratio $\rho_2/\rho_1$. In Figure \ref{fig:cohesion_density}, we plot the normalized cohesion as a function of the asphericity parameter for a constant $D_2/D_1=0.7$, varying the density ratio. We see similar trends between prolate and oblate secondaries, but the trends are much stronger for prolate bodies. This demonstrates that for nearly spherical secondaries, the required cohesion decreases as density ratio increases. Conversely, for more aspherical secondaries, the required cohesion increases with increasing density ratio.

\subsection{Spheres} \label{sec:spheres}

\begin{figure}[ht!]
   \centering
   \includegraphics[width = 3in]{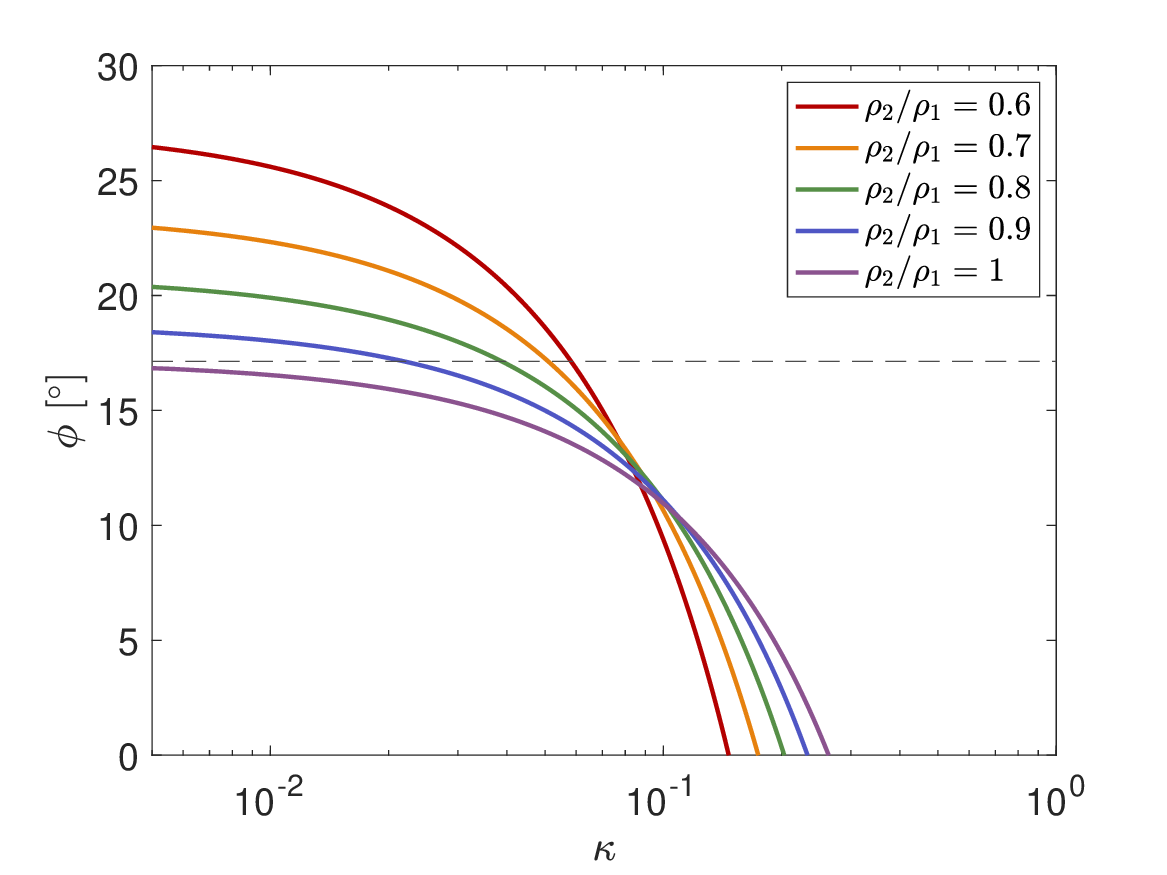} 
   \caption{The minimum friction angle as a function of normalized cohesion for the case of two equal-sized spheres. We show results for various density ratios, and the solution for the cohesionless equal mass case is plotted as a dashed line.}
   \label{fig:spheres}
\end{figure}

Next we examine the special case of two equal-sized spheres forming a contact binary, making the geometry of the problem much simpler. For the shapes, we have $a_1=b_1=c_1=a_2=b_2=c_2=R$. Under the equal mass case, this problem simplifies even further. Using this assumption we can directly solve for the friction angle required for cohesionless spheres:
\begin{equation}
    \phi = \sin^{-1}\left(\frac{3\sqrt{19}}{40+\sqrt{19}}\right)
\end{equation}
which gives a solution of $\phi\approx17.14^\circ$. Note this solution is independent of the size or mass of the spheres, as long as the sizes and masses are equal between the two spheres. As we vary the cohesion and density ratio, we obtain the curves shown in Figure \ref{fig:spheres}, which plots the friction angle as a function of normalized cohesion. Notably, this demonstrates that for a nominal friction angle of $30^\circ$, no cohesion is required for the equal-sized sphere case for reasonable density ratios. Figure \ref{fig:spheres} also shows the asymptote for the equal mass case as a dashed line where $\phi\approx17.14^\circ$.

Relaxing the constraint of equal-sized spheres sheds more light on the physical parameters that allow a contact binary to form. This approach allows us to investigate how the size ratio affects the required friction and cohesion in the body, without the influence of the shape.

\begin{figure}[ht!]
    \gridline{
            \fig{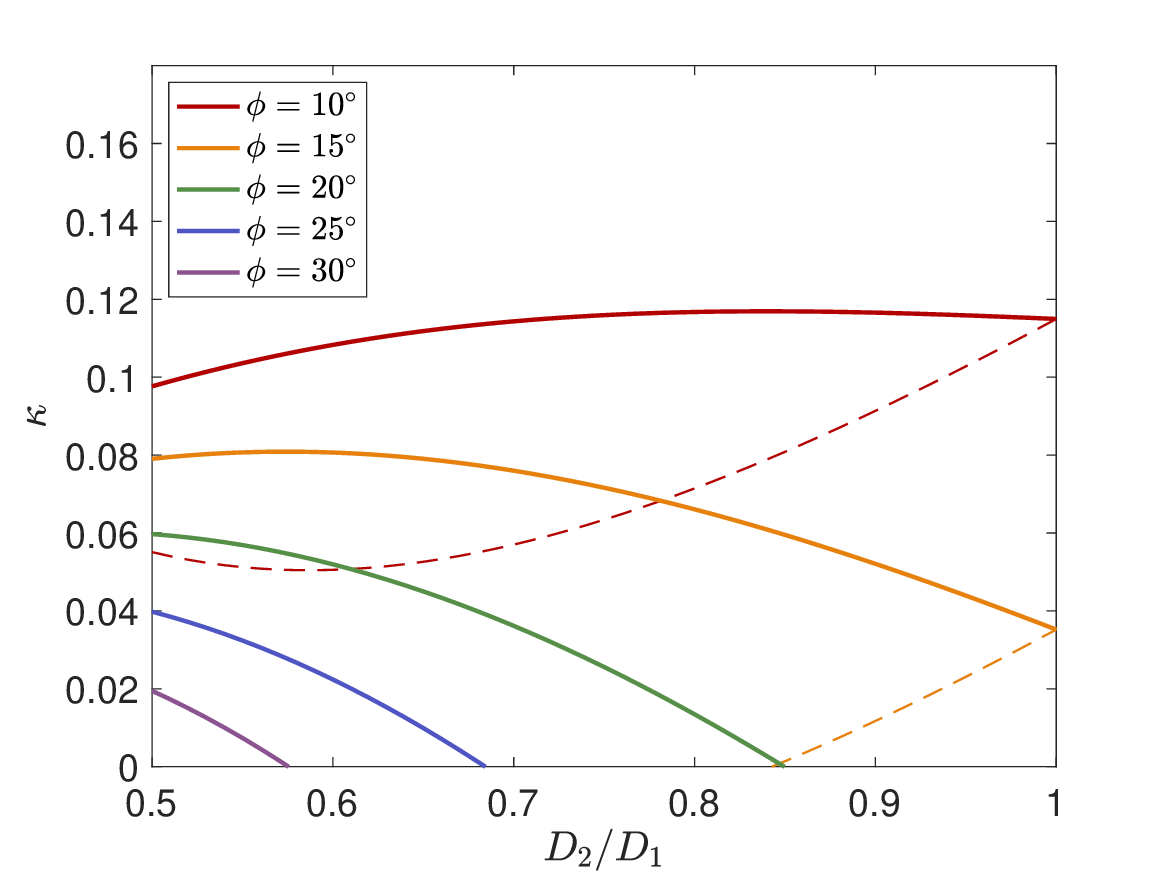}{0.45\textwidth}{(a) Required cohesion as a function of size ratio.}
            \fig{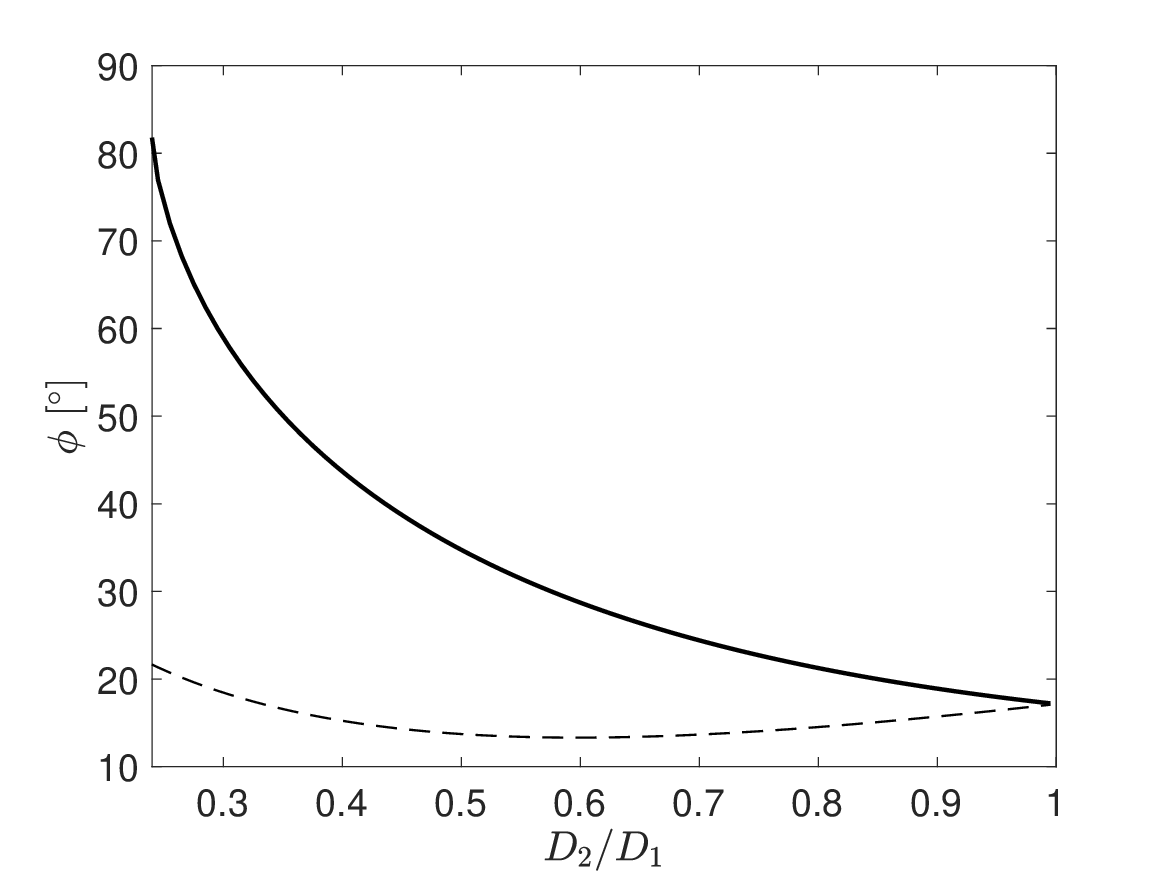}{0.45\textwidth}{(b) Required friction angle for cohesionless spheres.}
            }
    \caption{Plots showing the required cohesion and friction as a function of the size ratio of two spheres with equal densities. (a) shows the cohesion for various friction angles and (b) shows the required friction for cohesionless spheres. The solid line corresponds to the secondary and the dashed line to the primary,}
    \label{fig:spheres2}
\end{figure}

Figure \ref{fig:spheres2} plots the cohesion and friction required for two spherical bodies. We also plot the cohesion and friction required in the primary as dashed lines. For small friction angles, the cohesion required in the primary is less than that in the secondary. For larger friction angles, there is no cohesion required in the primary. The higher required cohesion in the secondary means the secondary will always fail before the primary, justifying our earlier assumption. Furthermore, for friction angles larger than $30^\circ$, no cohesion is required in either the primary or secondary for the observed size ratios.

This also shows the friction required when there is no cohesion, demonstrating that the minimum friction angle decreases with increasing size ratio for cohesionless spheres, up to the point of equal size bodies where it is equal to $17.14^\circ$ as calculated previously. In the cohesionless case, the friction required in the secondary is always greater than that for the primary, again showing the secondary is more susceptible to failure.

\section{In the Context of Asteroid Populations} \label{sec:population}

Next, we compare our population of contact binaries with other similar populations, namely binary asteroids and asteroid pairs. Binary asteroids are any systems in which the secondary is in orbit around the primary, whereas asteroid pairs are two asteroids with similar heliocentric orbits that share a common origin. We take our binary asteroid data from \cite{pravec2016binary,pravecdatabase}, and our asteroid pair data from \cite{pravec2019asteroid}.

The asteroid pair data is reported as lightcurve magnitudes and amplitudes. We convert these to diameter ratios and elongations using simple relationships, which we summarize in Appendix \ref{apdx:C}. In calculating the diameter ratios we assume both asteroids have the same albedo, and in calculating the shape elongation we assume we are viewing from the asteroid's equator. 

\begin{figure}[ht!]
    \gridline{
            \fig{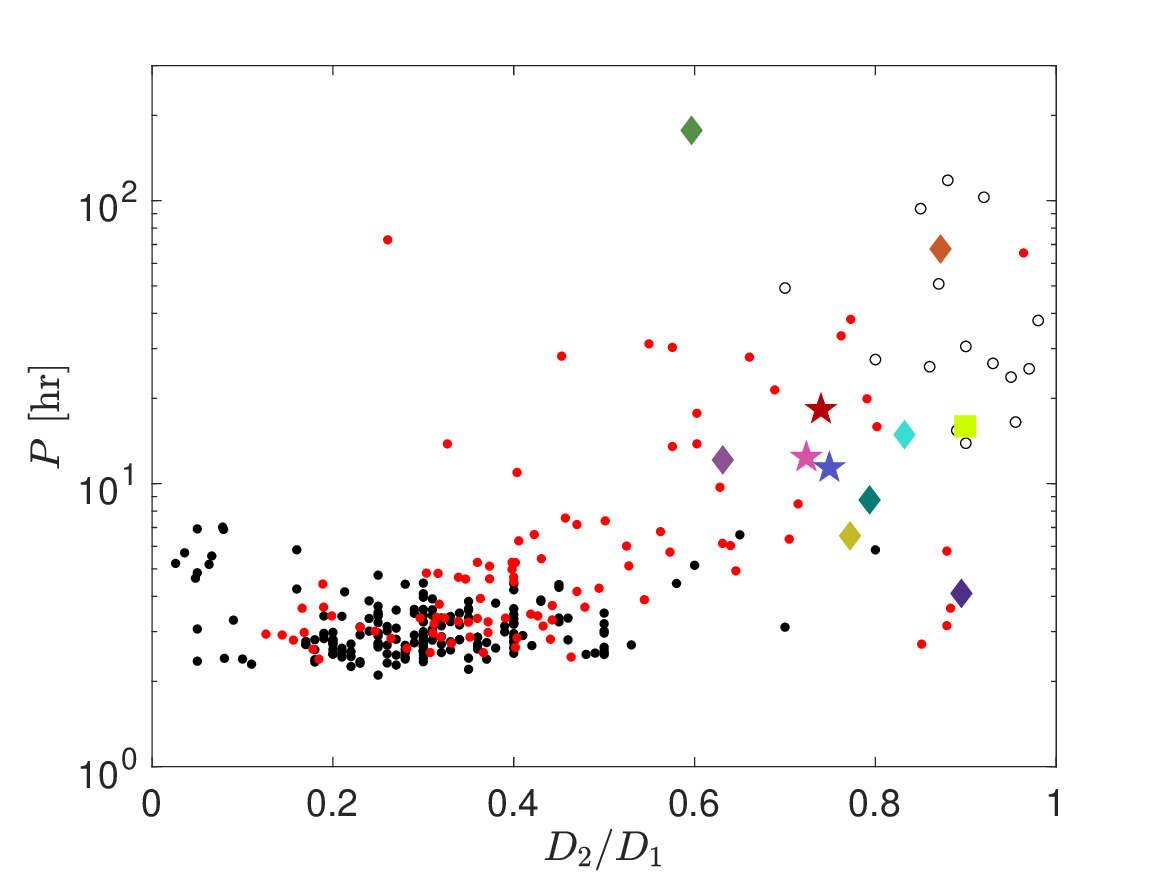}{0.45\textwidth}{(a) Rotation period as a function of size ratio.}
            \fig{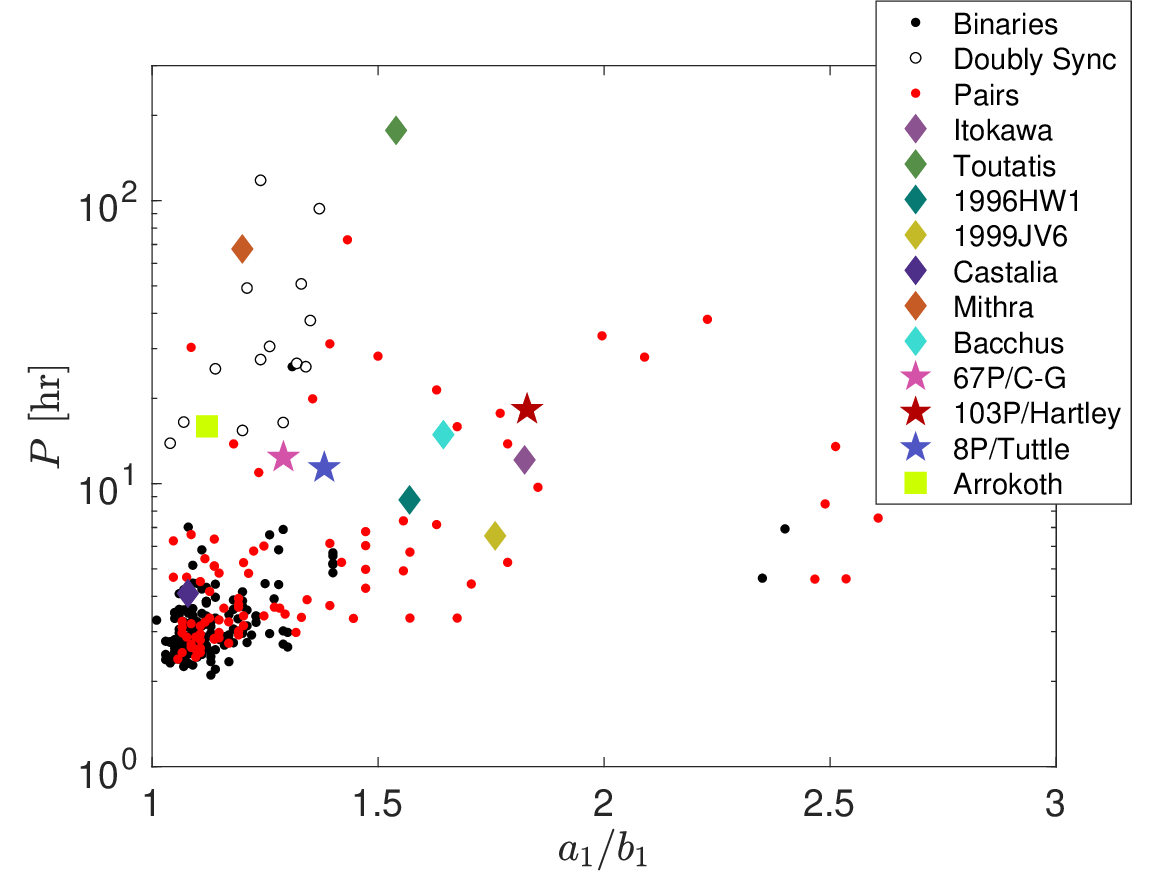}{0.45\textwidth}{(b) Rotation period as a function of primary elongation.}
            }
    \caption{Plots showing the populations of binary asteroids, asteroid pairs, and contact binaries, plotting (a) the primary's rotation period as a function of size ratio and (b) the primary's rotaiton period as a function of the primary's elongation. Doubly synchronous binary asteroids are plotted separately as an empty black circle.}
    \label{fig:population}
\end{figure}

Figure \ref{fig:population} shows the rotation period of the primary as a function of both the shape ratio and primary elongation. In these plots we show the populations of binary asteroids, asteroid pairs, and the contact binaries studied in this work. \cite{pravec2010formation} already showed how asteroid pairs converge to a similar distribution as binary asteroids at small sizes, and \cite{pravec2019asteroid} pointed out several asteroid pairs have a binary asteroid as one member. We add to this comparison the 11 contact binaries studied in this work.

From Figure \ref{fig:population}, we can identify some interesting trends. All three populations follow a similar distribution between rotation period and size ratio. For asteroid pairs, this means the primary and secondary are formed from the same rapidly-rotating progenitor \citep{pravec2010formation}. However, the same trend in binary asteroids is driven by tidal synchronization of the primary in doubly-synchronous binaries through tidal evolution. We see this with doubly-synchronous binaries having slow primary rotations and size ratios, and singly-synchronous binaries having fast primary rotations and small size ratios.

We also see none of the contact binaries studied in this work and none of the doubly synchronous binary asteroids has a size ratio smaller than about 0.6. This approximately corresponds to the critical mass ratio of 0.2, for which the secondary component can escape if the system is spun up to the point of fission, identified by \cite{pravec2010formation} for asteroid pairs, and also shown by \cite{hirabayashi2016fission} for bilobate comets.

In plotting the primary elongation, we start to see divergence between these populations. Generally, binary asteroids have relatively low elongations, whereas asteroid pairs and contact binaries see more elongated primaries. Overall from these plots, we see contact binaries share similarities with asteroid pairs and doubly-synchronous binary asteroids, but not singly-synchronous binary asteroids.

\section{Discussion} \label{sec:discussion}
For contact binary asteroids to form, the secondary must survive the merger with the primary. This means the Roche limit of the primary must be within the sum of the radii of the primary and secondary. Applying this criteria to several contact binaries, we have estimated the bulk cohesion required in these bodies. While the cohesion also depends on the bodies' friction angle, in general we find on the order of 1-100 Pa is required to form a contact binary in these systems. This is an order of magnitude study, so we do not have formal uncertainties on our estimates. But unless these bodies have a high friction angle, at least some amount of cohesion is required to survive the merging process.

These cohesion values are consistent with observations. On the low end, Bennu and Ryugu have surfaces on the order of 1 Pa cohesion \citep{walsh2022near,jutzi2022constraining}. Other observations show small rubble piles having global strengths of 10s to 100s Pa cohesion \citep{rozitis2014cohesive,hirabayashi2014constraints,jewitt2017anatomy}. Therefore the minimum cohesion values calculated in this work are physically plausible.

The required cohesion strongly depends on the shape and relative size of the secondary. The required strength is expected to vary with the shape of the body, as it may pass closer to a natural equilibrium configuration for its given spin rate as the shape is varied \citep{washabaugh2002energy}. For a given size ratio, the required cohesion increases as the secondary becomes more aspherical. This trend is stronger for prolate secondaries than oblate. As we increase size ratio, for nearly spherical secondaries the required cohesion increases then decreases, whereas for more aspherical secondaries the required cohesion only increases with increasing size ratio. Thus, there is some critical shape where the relationship between cohesion and size ratio changes. Around this critical value, cohesion is nearly independent of size ratio for large size ratios. For small size ratios, cohesion always increases with increasing size ratio regardless of secondary shape.

The density ratio between the secondary and primary also affects the required cohesion. For nearly spherical shapes, increasing the density of the secondary relative to the primary requires less cohesion. On the contrary, for more aspherical shapes increasing the density ratio requires more cohesion. Again, this effect is more pronounced for prolate secondaries than oblate. Thus, the dependence on density ratio also flips at the critical shape.

This analysis applied to two equal-sized spheres reveals that cohesionless spheres only require around a $17^\circ$ friction angle for equal bulk densities. As the density ratio diverges from unity, the required friction increases, but for all reasonable density ratios, no cohesion is required for a $30^\circ$ friction angle. This carries implications for Selam, the secondary of (152830) Dinkinesh, which appears to be two roughly equally-sized spheres \citep{dinkinesh}. While the formation mechanism here is likely different as this is a contact binary within a binary asteroid, this still suggests very little, if any, cohesion is needed for the merger of two progenitor satellites.

Of particular interest is placing our contact binaries in the context of the population of multi-body systems, specifically binary asteroids and asteroid pairs. Of the contact binaries examined in this work, none have a size ratio smaller than 0.6, which corresponds to the critical mass ratio for a fissioned asteroid (assuming identical homogeneous densities) of 0.2 identified by \cite{scheeres2009stability}. Fissioned systems below this mass ratio are dynamically unstable and form pairs \citep{jacobson2011dynamics}. We also observe these contact binaries have rotation periods similar to doubly-synchronous binaries and asteroid pairs of similar size ratios. 

It is worth noting that all three of these populations may be in different phases of their evolution. Binary asteroids can tidally evolve, causing a decrease in the primary's rotation, as we see in Figure \ref{fig:population}. While this may be counteracted by YORP, these systems are nonetheless dynamically evolved since their formation. However, small size-ratio binary asteroids are probably less dynamically evolved owing to weaker tidal forces within the system. At these small size ratios, the populations of binary asteroids and asteroid pairs converge. A similar argument is made for contact binaries, as the low-speed collision between the two lobes can also dynamically evolve the system. YORP evolution also affects contact binaries; for example Toutatis is currently in a non-principal axis rotation state \citep{hudson1995shape} and is somewhat of an outlier with its slow rotation period. However, despite their dynamical evolution, the three populations share similar features in the distributions investigated in this work. Taken in the context of the literature, this suggests all three populations are different evolutionary paths taken by rubble pile bodies. Indeed, members could switch between these populations, for example the secondary escaping a binary asteroid to become an asteroid pair, the secondary in a doublys-synchronous binary asteroid colliding with the primary to form a contact binary, or a contact binary spinning up to the point of fission and becoming a doubly-synchronous binary asteroid or an asteroid pair.

Interestingly, we see the shapes of the primaries of contact binaries are usually elongated. Looking at the populations of binary asteroids and asteroid pairs, we see binary asteroids usually have more equatorially symmetric primaries, whereas asteroid pairs can have more elongated primaries. Thus, in terms of primary shape, our population of contact binaries seems more consistent with asteroid pairs than binary asteroids.

While there is general consensus that multi-body systems are formed by spin-up to the point of mass loss, the actual mechanism of mass loss is unclear. Generally there are two hypotheses: either mass shedding from the equator \citep{walsh2008rotational,walsh2012spin,madeira2023dynamical,agrusa2024direct} or critical failure resulting in fission of the parent body \citep{jacobson2011dynamics}. These failure modes were characterized in more detail by \cite{zhang2022inferring}, who showed that failure by mass shedding occurs for bodies with high friction and no cohesion, whereas failure by fission occurs for bodies with moderate friction and some cohesion. Note both these scenarios are consistent with the required cohesion and friction found in this work for the formation of a contact binary. In their analyses, \cite{walsh2012spin} and \cite{zhang2018rotational} showed failure and satellite formation by mass shedding generally produces equatorially symmetric primaries, meaning the formation of asteroid pairs is more consistent with failure by fission, as suggested by \cite{pravec2010formation}. The low elongations of binary asteroids suggest these systems are preferentially formed by mass shedding, but this should be the subject of further investigation. It is unclear how contact binaries fit into this picture. Given the similarities between the populations of asteroid pairs and contact binaries, it could be that contact binaries are failed asteroid pairs that do not separate, especially since none of the contact binaries investigated here have a mass ratio below the critical limit. Alternatively, given their similar distributions in shape and rotation, contact binaries could be collapsed doubly-synchronous binary asteroids. It's possible both mechanisms are at play in the formation of these objects. This undoubtedly requires further analysis, ideally with information about more contact binaries. Nevertheless, this indicates both tensile failure and mass shedding could be working in parallel to evolve the population of small rubble piles. One exception is Arrokoth, which may have been formed by the streaming instability rather than rotational failure \citep{nesvorny2010formation}. However, we note that while Arrokoth does have the lowest primary elongation, it still generally fits the population of the other contact binaries studied here.

This study is only meant to be an order of magnitude analysis, but the major source of uncertainty comes from the ellipsoidal fit to the bodies, as the results are sensitive to shape. Additionally, this means density inhomogeneities could have a significant impact on these results. Another potential shortcoming is the failure model used, as this model assumes the primary is large and spherical. However, as we have shown in our results, this is not always the case. The primary is frequently elongated in contact binaries, and some of the systems investigated here are near equal mass. The current approach still serves to provide an estimate for the minimum cohesion needed to keep these contact binaries intact. Future analylsis should reformulate the theory when both components are non-spherical.

In this work, we have analyzed the hypothesis that contact binaries are formed by two energetically bound progenitor bodies that experience a collapse of their mutual orbit. We have found this formation mechanism only requires reasonably small amounts of cohesion and friction within a rubble pile body. This suggests this hypothesis is physically feasible, and does not require any large monolithic components within the bodies to survive the merger process.

\appendix
\section{Ellipsoid Fitting} \label{apdx:A}
\begin{figure}[ht!]
   \centering
   \includegraphics[width = 3in]{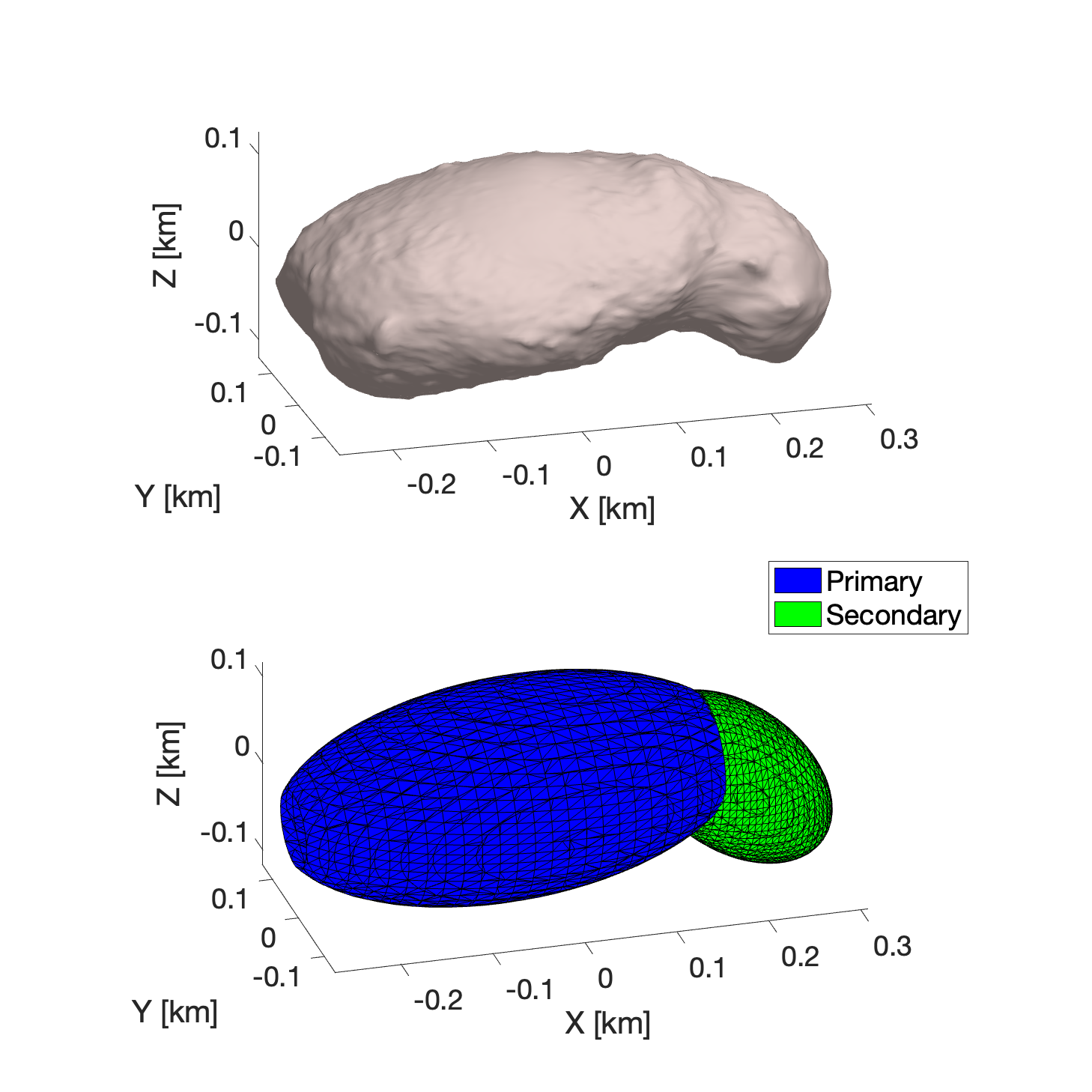} 
   \caption{The polyhedral shape model (top) and the ellipsoidal model (bottom) for Itokawa.}
   \label{fig:itokawa_model}
\end{figure}

\begin{figure}[ht!]
   \centering
   \includegraphics[width = 3in]{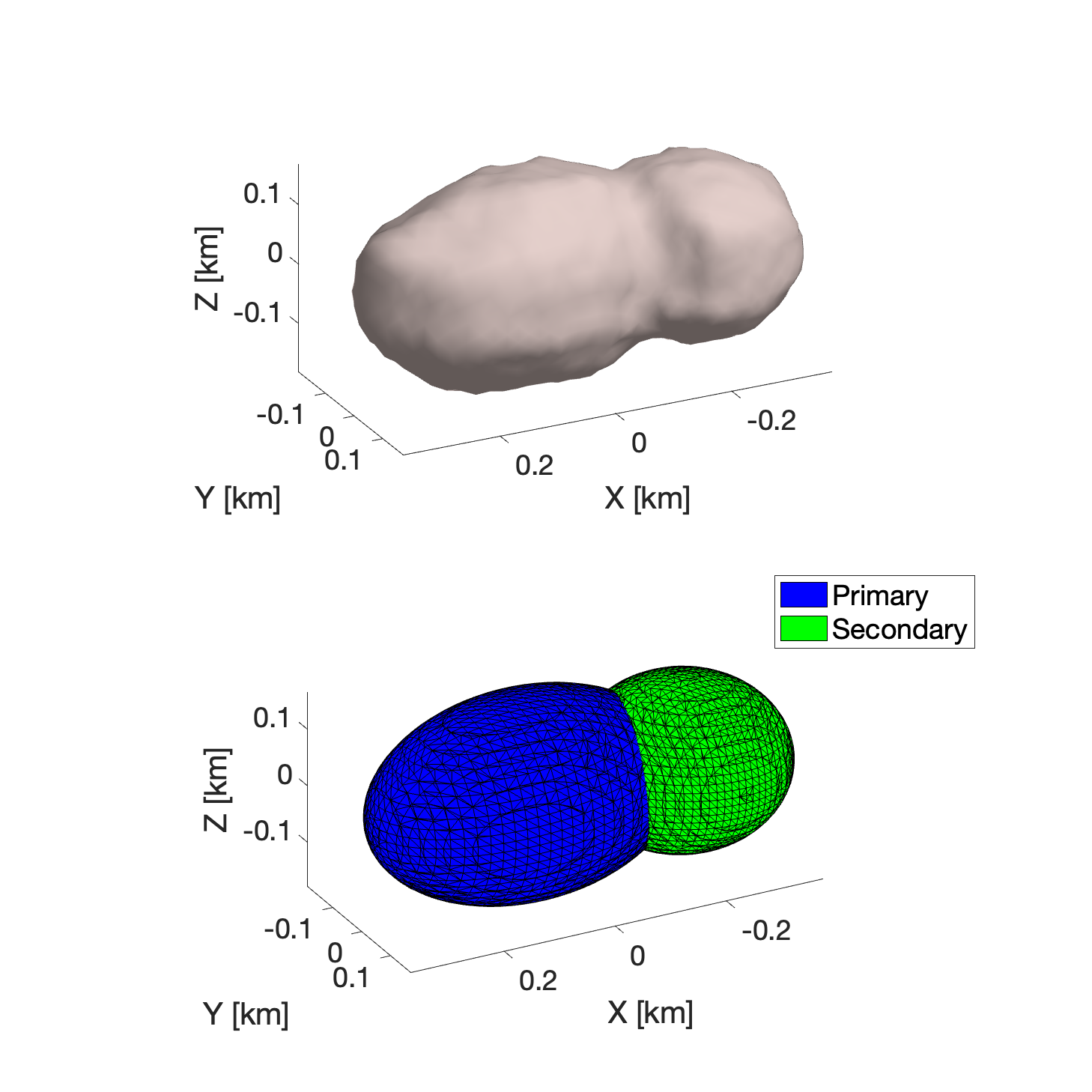} 
   \caption{The polyhedral shape model (top) and the ellipsoidal model (bottom) for 1999 JV6.}
   \label{fig:1999jv6_model}
\end{figure}

Here we show a few representative cases of our ellipsoidal fitting results. In these figures, we compare the polyhedral shape model with the ellipsoid fit for the lobes of the contact binary. Figure \ref{fig:itokawa_model} shows the comparison for Itokawa, while \ref{fig:1999jv6_model} shows the comparison for 1999 JV6. This illustrates good agreement between the actual shape model and our ellipsoidal approximation.

\section{Strength Model} \label{apdx:B}
Here we describe the failure model used in this work, as derived in \cite{holsapple2008tidal,holsapple2006tidal}. This theory calculates the breakup conditions for homogeneous ellipsoidal bodies subject to tidal forces from a larger primary, and is appropriate for granular rubble-pile bodies. We use the Drucker-Prager failure criteria:
\begin{equation}
    \sqrt{J_2}\leq k-sI_1
\end{equation}
where $I_1$ is the first invariant of the Cauchy stress tensor, defined as its trace:
\begin{equation}
    I_1 = \sigma_x + \sigma_y + \sigma_z
\end{equation}
and $J_2$ is the second invariant of the deviatoric stress tensor, defined as
\begin{equation}
    J_2 = \frac{1}{6}\left[(\sigma_x-\sigma_y)^2+(\sigma_y-\sigma_z)^2+(\sigma_z-\sigma_x)^2+\tau^2_{xy}+\tau^2_{yz}+\tau^2_{zx}\right].
\end{equation}
Here, $\sigma_x$, $\sigma_y$, and $\sigma_z$ are the normal stresses and $\tau_{xy}$, $\tau_{yz}$, and $\tau_{zx}$ are the shear stresses in the $x$, $y$, and $z$ directions. In our failure criteria we also have $k$, which is the bulk macroscopic cohesion of the body, and $s$, which represents the shear strength and is related to the friction angle $\phi$ by
\begin{equation}
    s = \frac{2\sin\phi}{\sqrt{3}(3-\sin\phi)}.
\end{equation}
In the principal stress state, the shear stresses are identically zero. The principal normal stresses caused by rotation and tidal gravity, once averaged over the ellipsoid, are
\begin{equation}
\begin{aligned}
    \bar{\sigma}_1 &= \frac{a_2^2}{5}\left(\rho_2\omega^2-2\pi\rho_2^2GA_x+\frac{8\pi}{3}G\rho_2\rho_1\left(\frac{2d}{D_1}\right)^{-3}\right)\\
    \bar{\sigma}_2 &= \frac{b_2^2}{5}\left(\rho_2\omega^2-2\pi\rho_2^2GA_y-\frac{4\pi}{3}G\rho_2\rho_1\left(\frac{2d}{D_1}\right)^{-3}\right)\\
    \bar{\sigma}_3 &= \frac{c_2^2}{5}\left(-2\pi\rho_2^2GA_z-\frac{4\pi}{3}G\rho_2\rho_1\left(\frac{2d}{D_1}\right)^{-3}\right).
\end{aligned}
\end{equation}
If the secondary is tidally locked to the primary, the spin rate is calculated as
\begin{equation}
    \omega^2 = \frac{G(m_1+m_2)}{d^3}.
\end{equation}
For a contact binary, we set the separation between the bodies, $d$, equal to the sum of the two radii, which is appropriate for our averaged approach:
\begin{equation}
    d = \frac{D_1+D_2}{2}.
\end{equation}
The $A_i$ terms terms are defined as
\begin{equation}
\begin{aligned}
    A_x &= \frac{b_2c_2}{a_2^2}\int_0^\infty{\frac{du}{(u+1)^{3/2}(u+(b_2/a_2))^{1/2}(u+(c_2/a_2))^{1/2}}}\\
    A_y &= \frac{b_2c_2}{a_2^2}\int_0^\infty{\frac{du}{(u+1)^{1/2}(u+(b_2/a_2))^{3/2}(u+(c_2/a_2))^{1/2}}}\\
    A_z &= \frac{b_2c_2}{a_2^2}\int_0^\infty{\frac{du}{(u+1)^{1/2}(u+(b_2/a_2))^{1/2}(u+(c_2/a_2))^{3/2}}}
\end{aligned}
\end{equation}
and are numerically calculated by quadrature.

Using the average normal stresses, we can calculate the cohesion and friction angle required to meet the Drucker-Prager failure criteria. 

\section{Lightcurve Analysis} \label{apdx:C}

Here we present the very simple lightcurve analysis equations used in this work. The diameter of an object given its absolute magnitude $H$ and albedo $p_v$ is \citep{warner2009asteroid}:
\begin{equation}
    \log{D}\mathrm{[km]}=3.1235-0.2H-0.5\log{p_v}.
\end{equation}
While we have the unknown albedo, this quantity cancels out when we calculate the diameter ratio between asteroid pairs, assuming the two asteroids have the same albedo. Thus:
\begin{equation}
    \frac{D_2}{D_1} = 10^{0.2(H_1-H_2)}.
\end{equation}

To calculate the elongation of the asteroid, we first have to know its spin pole. However, generally we do not have this information so we assume we are viewing the asteroid by its equator. This is calculated using the lightcurve amplitude $A$ as \citep{binzel1989asteroids}:
\begin{equation}
    \frac{a}{b} = 10^{0.4A}.
\end{equation}

\bibliography{bib}{}
\bibliographystyle{aasjournal}



\end{document}